\documentclass[traditabstract]{aa} 
\usepackage{amsmath}
\usepackage{graphicx}
\usepackage{txfonts}

\usepackage{natbib}
\bibpunct{(}{)}{;}{a}{}{,}
\usepackage{float}
\usepackage{calc}
\usepackage{textcomp}
\usepackage{placeins}
\usepackage{color}
\usepackage{rotating} 
\usepackage{lscape}
\usepackage{url}
\usepackage[colorlinks=true,linkcolor=blue,citecolor=blue]{hyperref}
\usepackage{subfig}
\usepackage[all]{draftcopy}
\usepackage{longtable}
\usepackage{array}
\usepackage{threeparttable}
\usepackage{lscape}
\usepackage{enumerate}
\DeclareTextSymbol{\degre}{OT1}{23}
\newcounter{savedfootnote}

\renewcommand{\epsilon}{\varepsilon} 
\def \microns{{\,$\mu$m}}

\begin{document}
\title{Constraining the recent star formation history of galaxies: an Approximate Bayesian Computation approach.}

\author{G.~Aufort\inst{1,2}, L.~Ciesla\inst{2}, P.~Pudlo\inst{1}, and V.~Buat\inst{3}.
}

\institute{	
Aix Marseille Universit\'e, CNRS, Centrale Marseille, I2M, Marseille, France
 \and
 Aix-Marseille  Universit\'e,  CNRS, LAM (Laboratoire d'Astrophysique de Marseille) UMR7326,  13388, Marseille, France
 \and
 Aix-Marseille  Universit\'e,  CNRS, LAM (Laboratoire d'Astrophysique de Marseille) UMR7326, Institut Universitaire de France (IUF),  13388, Marseille, France
}	

   \date{Received; accepted}

  \abstract
{
Although galaxies are found to follow a tight relation between their star formation rate and stellar mass, they are expected to exhibit complex star formation histories (SFH), with short-term fluctuations.
The goal of this pilot study is to present a method that will identify galaxies that are undergoing a strong variation of star formation activity in the last tens to hundreds Myr. 
In other words, the proposed method will determine whether a variation in the last few hundreds of Myr of the SFH is needed to properly model the SED rather than a smooth normal SFH.
To do so, we analyze a sample of COSMOS galaxies with $0.5<z<1$ and $\log M_*>8.5$ using high signal-to-noise ratio broad band photometry.
We apply Approximate Bayesian Computation, a state-of-the-art statistical method to perform model choice, associated to machine learning algorithms to provide the probability that a flexible SFH is preferred based on the observed flux density ratios of galaxies.
We present the method and test it on a sample of simulated SEDs.
The input information fed to the algorithm is a set of broadband UV to NIR (rest-frame) flux ratios for each galaxy.
The choice of using colors is made to remove any difficulty linked to normalization when using classification algorithms.
The method has an error rate of 21$\%$ in recovering the right SFH and is sensitive to SFR variations larger than 1\,dex.
A more traditional SED fitting method using CIGALE is tested to achieve the same goal, based on fits comparisons through Bayesian Information Criterion but the best error rate obtained is higher, 28$\%$.
We apply our new method to the COSMOS galaxies sample.
The stellar mass distribution of galaxies with a strong to decisive evidence against the smooth delayed-$\tau$ SFH peaks at lower M$_*$ compared to galaxies where the smooth delayed-$\tau$ SFH is preferred. 
We discuss the fact that this result does not come from any bias due to our training.
Finally, we argue that flexible SFHs are needed to be able to cover that largest SFR-M$_*$ parameter space possible. 
}

   \keywords{Galaxies: evolution, fundamental parameters}

   \authorrunning{Aufort et al.}
   \titlerunning{ABC applied to SFH}

   \maketitle

\section{\label{intro}Introduction}

The tight relation linking the star formation rate (SFR) and stellar mass of star-forming galaxies,the so-called main sequence (MS), opened a new window in our understanding of galaxy evolution \citep{Elbaz07,Noeske07}.
It implies that the majority of galaxies are likely to form the bulk of their stars through steady-state processes rather than violent episodes of star formation.
However, this relation has a scatter of $\sim$0.3\,dex \citep{Schreiber15} that is found to be relatively constant at all masses and over cosmic time \citep{Guo13,Ilbert15,Schreiber15}.
One possible explanation of this scatter could be its artificial creation by the accumulation of errors in the extraction of photometric measurements and/or in the determination of the SFR and stellar mass in relation with model uncertainties. 
However, several studies have found a coherent variation of physical galaxy properties such as the gas fraction \citep{Magdis12}, Sersic index and effective radius \citep{Wuyts11}, and U-V color \citep[e.g.,][]{Salmi12}, suggesting that the scatter is more related to the physics than to measurement and model uncertainties. 
Furthermore, oscillations of the SFR resulting from a varying infall rate and compaction of star-formation have been proposed to explain the MS scatter \citep{Sargent14,Scoville16,Tacchella16} and even suggested by some simulations \citep[e.g.,][]{DekelBurkert14}.

To decipher if the scatter is indeed due to star formation history (SFH) variations, one must be able to put constraint on the recent star formation history (SFH) of galaxies, to reconstruct their path along the MS.
This information is embedded in the spectral energy distribution (SED) of galaxies.
However, recovering it through SED modeling is complex and subject to many uncertainties and degeneracies.
Indeed, galaxies are expected to exhibit complex SFHs, with short-term fluctuations, requiring sophisticated SFH parametrizations to model them \citep[e.g.,][]{Lee10,Pacifici13,Behroozi13,Pacifici16,Leja19}.
The implementation of these models is complex and large libraries are needed to model all galaxies properties. Numerous studies have, instead, used simple analytical forms to model galaxies SFH \citep[e.g.,][]{Papovich01,Maraston10, Pforr12,Gladders13,Simha14,Buat14,Boquien14,Ciesla15,Abramson16,Ciesla16,Ciesla17}.
However, SFH parameters are known to be difficult to constrain from broadband SED modeling \citep[e.g.,][]{Maraston10,Pforr12,Buat14,Ciesla15,Ciesla17,Carnall19}.

\cite{Ciesla16} and \cite{Boselli16} have shown on a sample of well-known local galaxies benefiting from a wealth of ancillary data, that a drastic and recent decrease of the star formation activity of galaxies can be probed as long as a good UV to NIR rest frame coverage is available. 
They showed that the intensity of the variation of SF activity can be relatively well constrained from broadband SED fitting.
Spectroscopy is however needed to bring information on the time when the change in star formation activity occurred \citep{Boselli16}.
These studies were made on well-known sources of the Virgo cluster, for which the quenching mechanism - ram pressure stripping - is known and HI observations allow a direct verification of the SED modeling results.
To go a step further, \cite{Ciesla18} have blindly applied the method on the GOODS-South sample aiming at identifying sources that underwent a recent and drastic decrease of their star-formation activity.
They compared the quality of the results from SED fitting using two different SFH and obtained a sample of galaxies where a modeled recent and strong decrease of SFR produced significantly better fits of the broad band photometry.
In this work, we aim at improving the method of \cite{Ciesla18} gaining in power by applying to a subsample of COSMOS galaxies a state-of-the-art statistical method to perform the SFH choice: the Approximate Bayesian Computation \citep[ABC, see, e.g. ][]{marin2012approximate,sisson2018handbook}.
Based on the observed SED of a galaxy, we want to choose the most appropriate SFH between a finite set. 
The main idea behind ABC is to rely on many simulated SEDs generated from all the SFHs in competition using parameters drawn from the prior.

The paper is organized as follows:
Sect.~\ref{astro} describes the astrophysical problem and presents the sample.
In Sect.~\ref{stat} we present the statistical approach as well as the results obtained from a catalog of simulated SEDs of COSMOS-like galaxies. In Sect.~\ref{synthetic} we compare the results of this new approach with more traditional SED modeling methods, and apply it to real COSMOS galaxies in Sect.~\ref{real}.
Our results are discussed in Sect.~\ref{conclusions}.

\section{\label{astro} Constraining the recent star formation history of galaxies using broad-band photometry}

\subsection{\label{c18}Building upon the method of \cite{Ciesla18}}
The main purpose of the study presented in \cite{Ciesla18} was to probe variations in SFH that occurred in very short timescales, i.e. on hundreds of Myrs.
Therefore, a large-number statistics was needed to be able to catch galaxies at the moment when these variations happened. 
They aimed at identifying galaxies that have recently undergone a rapid ($<$500\,Myr) and drastic downfall of their SFR (more than 80\%) from broadband SED modeling, since large photometric samples can provide the statistics needed to pinpoint these objects.

To perform their study, they took advantage of the versatility of the SED modeling code CIGALE\footnote{\url{https://cigale.lam.fr/}} \citep{Boquien19}.
CIGALE is a SED modeling software package that has two functions: a modeling function to create SEDs from a set of given parameters and a SED fitting function to derive the physical properties of galaxies from observations.   
Galaxies SEDs are computed from UV-to-radio taking into account the balance between the energy absorbed by dust in the UV-NIR and remitted in IR.
To build the SEDs, CIGALE uses a combination of modules including the star formation history assumption, either analytical, stochastic, or outputs from simulations \citep[e.g.,][]{Boquien14,Ciesla15,Ciesla17}, the stellar emission from stellar population models \citep{BruzualCharlot03,Maraston05}, the nebular lines, and the attenuation by dust \citep[e.g.,][]{Calzetti00,CharlotFall00}.

\begin{figure}[!h] 
  	\includegraphics[width=\columnwidth]{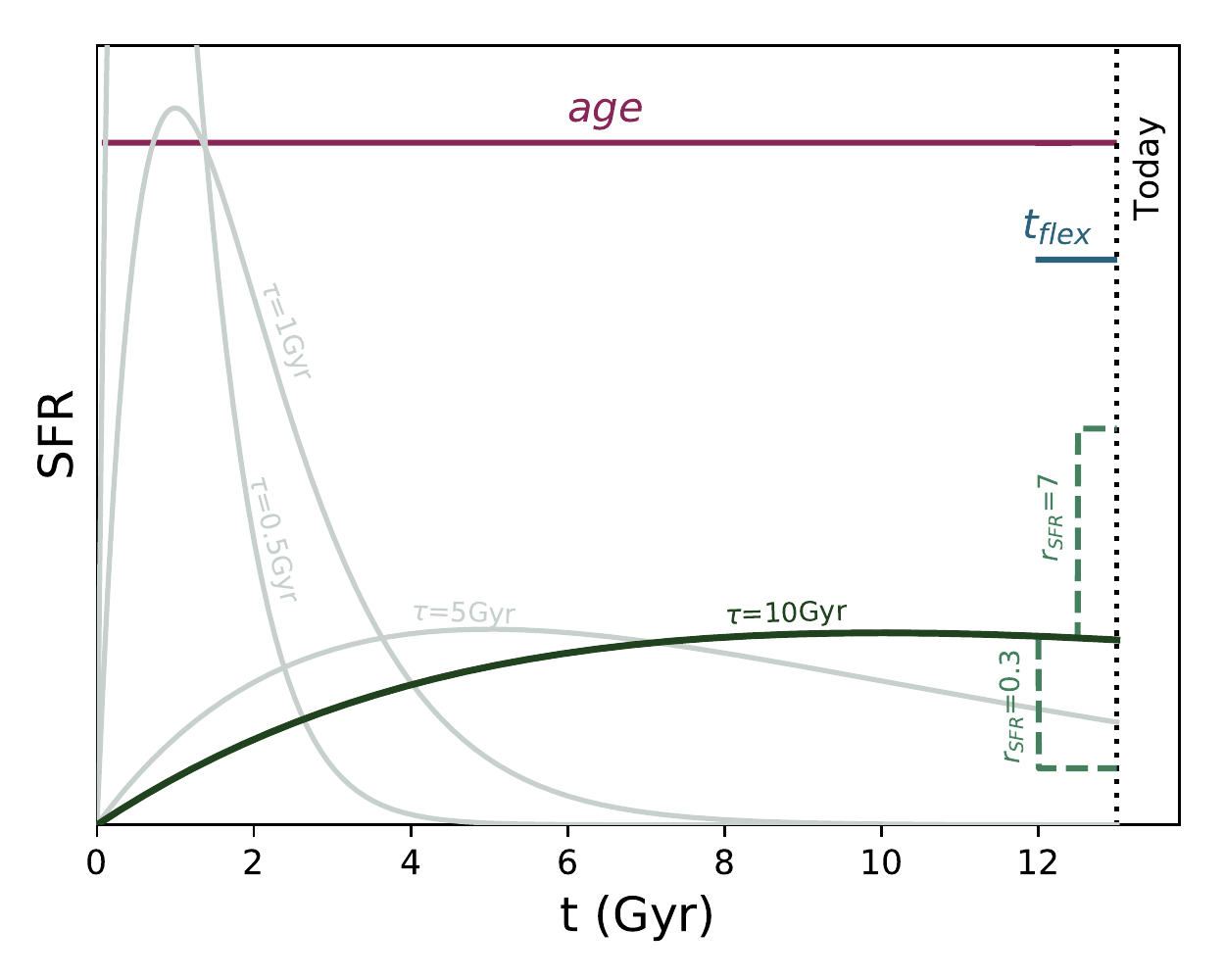}
  	\caption{\label{sfhschema}Examples of delayed-$\tau$ SFHs considered in this work (star formation rate as a function of cosmic time). Different SFHs using $\tau_{main}=$0.5, 1, 5, and 10\,Gyr are shown to illustrate the impact of this parameter (light green and dark green solid lines). An example of delayed-$\tau$ SFH with flexibility is shown in solid dark green with the flexibility in green dashed lines for ($age_{\mathrm{flex}}$=1\,Gyr \& $r_{\mathrm{SFR}}=$0.3) and  ($age_{\mathrm{flex}}$=0.5\,Gyr \& $r_{\mathrm{SFR}}=$7). }
\end{figure}

\cite{Ciesla18} compared the results of SED fitting on a sample of GOODS-South galaxies using two different SFHs: one normal delayed-$\tau$ SFH and one flexible SFH modeling a truncation of the SFH.
The normal delayed-$\tau$ SFH is given by the equation:
\begin{equation}        
\mathrm{SFR}(t) \propto t \times exp(-t/\tau_{main})\\
\end{equation} 
\noindent where SFR is the star formation rate, t the time, and $\tau_{main}$ is the e-folding time.
Examples of delayed-$\tau$ SFHs are shown in Fig.~\ref{sfhschema} for different values of $\tau_{main}$.
The flexible SFH is an extension of the delayed-$\tau$ model:
\begin{equation}        
\mathrm{SFR}(t) \propto
\begin{cases}
    t \times exp(-t/\tau_{main}), & \text{when}\ t\leq t_{flex} \\
    r_{\rm{SFR}} \times \mathrm{SFR}(t=t_{flex}), & \text{when}\ t>t_{flex} \\
\end{cases},
\end{equation}  
        
\noindent where $t_{flex}$ is the time at which the star formation is instantaneously affected, and $r_{\rm{SFR}}$ is the ratio between SFR$(t>t_{flex})$ and SFR$(t=t_{flex})$:
        
\begin{equation}
   r_{\rm{SFR}} = \frac{\mathrm{SFR}(t>t_{flex})}{\mathrm{SFR}(t_{flex})}.
\end{equation}

\noindent A representation of flexible SFHs is also shown in Fig.~\ref{sfhschema}.
The normal delayed-$\tau$ SFH is at first order a particular case of the flexible SFH for which $r_{\rm{SFR}}=1$.

To differentiate between the two models, \cite{Ciesla18} estimated the Bayesian Information Criterion (BIC, see Sect.~\ref{Bmc}) linked to the two models and put conservative limits on the difference between the two BIC to select the most suited model.
They showed that a handful of sources were better fitted using the flexible SFH, that assumes a recent instantaneous break in the SFH, compared to the more commonly used delayed-$\tau$ SFH.
In fact, they discussed that these galaxies have indeed physical properties that are different from the main population and characteristic of sources in transition.

The limited number of sources identified in the study of \cite{Ciesla18} (102 out of 6,680) was due to their will to be conservative in their approach and find a clean sample of sources that underwent a rapid quenching of star formation.
Indeed, they imposed that the instantaneous decrease of SFR was more than 80\% and that the BIC difference was larger than 10.
These criteria prevent a complete study of rapid variations in the SFH of galaxies as many of them would be missed. 
Furthermore, only decrease of SFR were considered and not the opposite, that is star formation bursts.
Finally, their method is time consuming as one has to run the CIGALE code twice, once per SFH model considered, to perform the analysis.
To go beyond this drawbacks and improve the method of \cite{Ciesla18}, we consider in the present pilot study a statistical approach, the Approximate Bayesian Computation, combined with classification algorithm to improve both the accuracy and the efficiency of their method.

\subsection{\label{sample}The sample}

In this pilot work, we use the wealth of data available on the COSMOS field.
The choice of this field is driven by the good spectral coverage of the data and the large statistics of sources available.

We draw a sample from the COSMOS catalog of \cite{Laigle16}.
A first cut is made to restrict ourselves to galaxies with a stellar mass \citep{Laigle16} higher than 10$^{8.5}$\,M$_{\odot}$.
Then, we restrict the sample to a relatively narrow range of redshift to minimize its impact on the SED and focus our method to the SFH effect on the SED. 
We thus select galaxies with redshift between 0.5 and 1, assuring sufficient statistics in our sample.
We use the broad bands of the COSMOS catalog, listed in Table~\ref{bands}. 
For galaxies with redshifts between 0.5 and 1, \textit{Spitzer}/IRAC3 probes the 2.9-3.9\microns\ wavelength range rest frame and \textit{Spitzer/}IRAC4 probes the 4-5.3\microns\ range rest frame. 
These wavelength ranges correspond to the transition between stellar and dust emission. 
To keep this pilot study simple we only consider the UV-to-NIR part of the spectrum, unaffected by dust emission. 

\begin{table}
	\centering
	\caption{COSMOS broad bands used in this work.}
	\begin{tabular}{l c c }
	 \hline\hline
	\textbf{Instrument} & \textbf{Band} & $\boldsymbol\lambda$ ($\boldsymbol\mu$\textbf{m}) \\ 
	\hline
	GALEX            & FUV & 0.153\\
	GALEX            & NUV & 0.229 \\
	CFHT        & $u'$ & 0.355 \\
	SUBARU          & B & 0.443\\
	SUBARU         & V & 0.544\\	   
	SUBARU         & r & 0.622\\	 
	Suprime Cam    & $i'$ & 0.767\\	
	Suprime Cam    & $z'$ & 0.902\\	
	VISTA            & Y & 1.019\\	
	VISTA            & J & 1.250\\	
	VISTA            & H & 1.639\\	
	VISTA            & Ks & 2.142\\	
	\textit{Spitzer} & IRAC1 & 3.6\\	
	\textit{Spitzer} & IRAC2 & 4.5\\	
	\hline
	\label{bands}
	\end{tabular}
\end{table}

One aspect of the ABC method that is still to be developed is how to handle missing data.
In our astrophysical application, we identify several types of missing data.
First there is the impact of redshifting that is the fact that a galaxy is undetected at wavelength shorter than the Lyman break at its redshift.
Here, the absence of detection provides an information on the galaxy coded in its SED.
Another type of missing data is linked to the definition of the photometric surveys: the spatial coverage is not exactly the same in every bands and the different sensitivity limits yields to undetected galaxies due to the faintness of their fluxes.
To keep the statistical problem simple in this pilot study, we remove galaxies that are not detected in all bands.
This strong choice is motivated by the fact that the ABC method that we use in this pilot study has not been tested and calibrated in the case of missing data such as extragalactic field surveys can produce.
The impact of missing data on this method would require an important work of statistical research which is beyond the scope of this paper.

\begin{figure}[!h] 
  	\includegraphics[width=\columnwidth]{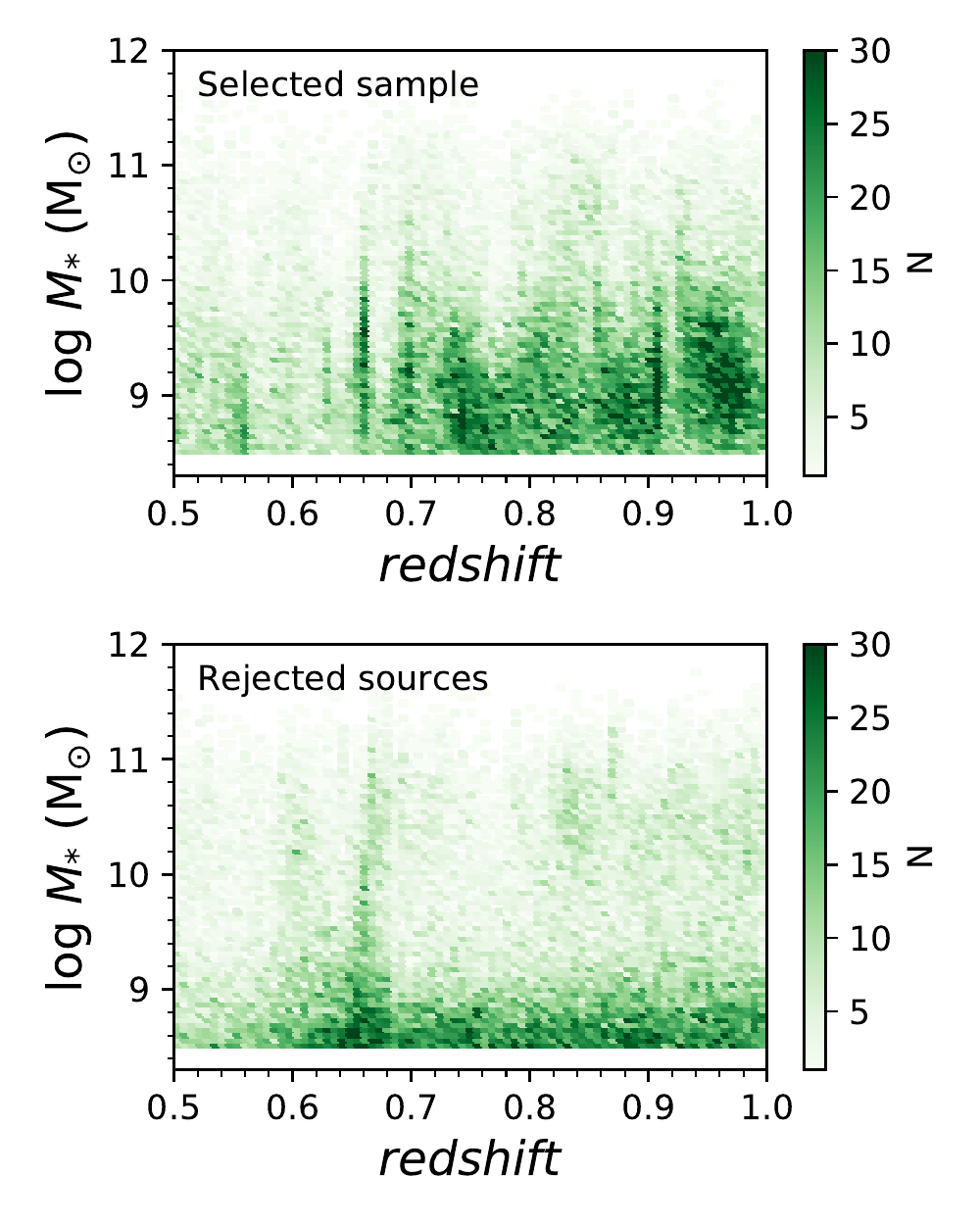}
  	\caption{\label{mass_vs_z}Stellar mass from \cite{Laigle16} as a function of redshift for the final sample (\textbf{top panel}) and for the rejected galaxies following our criteria (\textbf{bottom panel}).}
\end{figure}

As an additional constraint, we select galaxies with a SNR equal or greater than 10.
However, given the importance of the NUV band \citep{Ciesla16,Ciesla18} and the faintness of the fluxes compared to the other bands, we \textbf{relax} our criteria to a SNR of 5 for this band.
The first motivation for this cut is again to keep our pilot study simple, but we show in Appendix~\ref{apImp} that indeed this SNR cut is relevant.
In the following, we will consider a final sample composed of 12,380 galaxies for which the stellar mass distribution as a function of redshift is shown in Fig.~\ref{mass_vs_z} (top panel) and the distribution of the rejected sources in the bottom panel of the same figure.

\begin{figure}[!h] 
  	\includegraphics[width=\columnwidth]{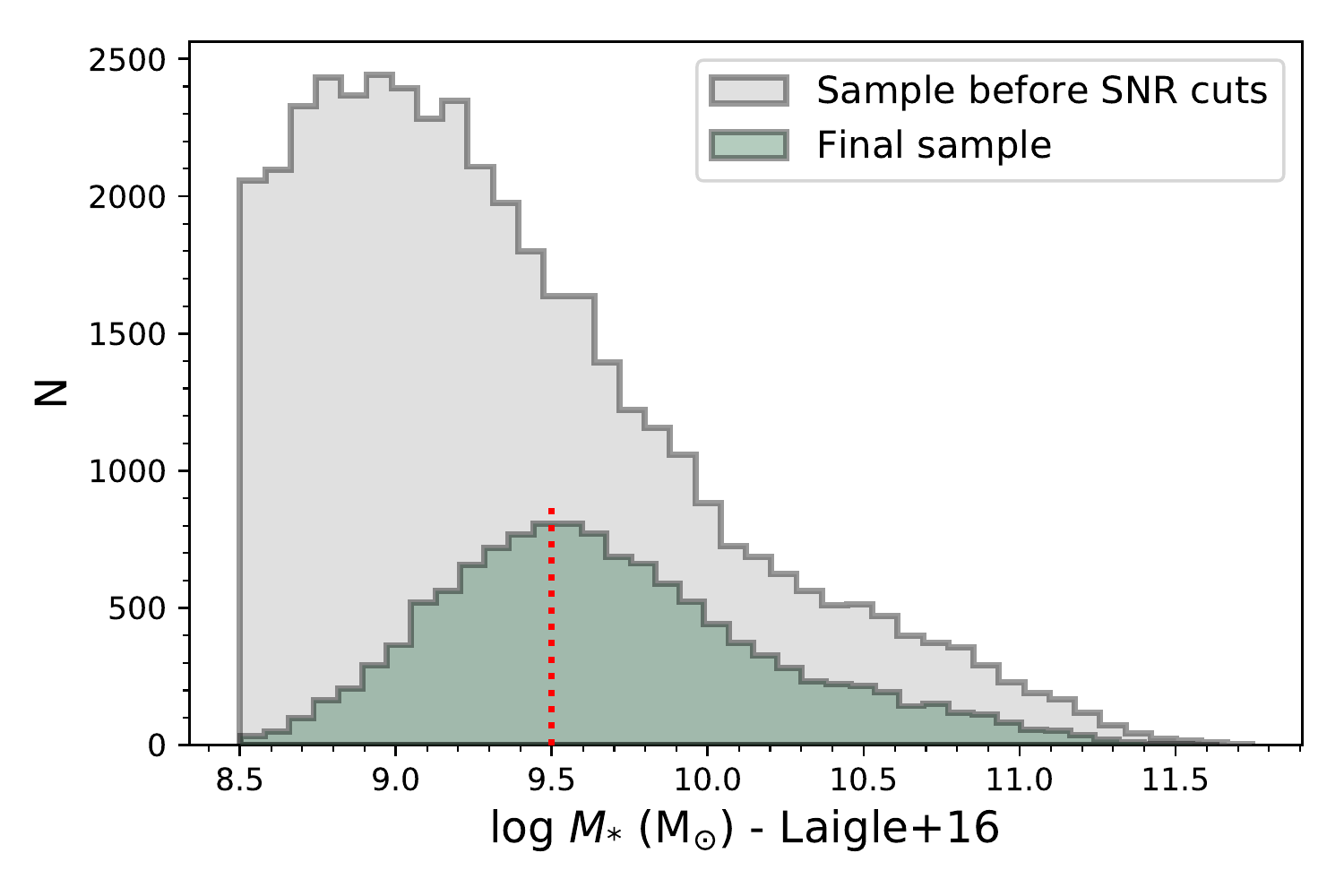}
  	\caption{\label{mass_comp}Distribution of stellar mass for the sample before the SNR cut (grey) and the final sample (green). The red dotted line indicated the limit above which our final sample is considered as complete. The stellar masses indicated here are from \cite{Laigle16}.}
\end{figure}

The stellar mass distribution, from \cite{Laigle16}, of the final sample is shown in Fig.~\ref{mass_comp}.
As a sanity check, we verify that above 10$^{9.5}$\,M$_{\odot}$, the stellar mass, star formation rate, and specific star formation rate distributions are similar.
Our selection criteria mostly affect low mass galaxies which is expected since we made SNR cuts.

Given the wide ranges of redshift, stellar masses, and SED shapes considered in our study, there is a normalization aspect that needs to be taken into account.
Indeed, this diversity in galaxies' properties translates into a large distribution of fluxes in a given photometric band, spanning over several orders of magnitude: 8 orders of magnitudes in the FUV band and 6 in the Ks band, for instance.
This parameter space is very challenging for classification algorithms.
To avoid this problem, we compute flux ratios.
First we combine each flux with the closest one in terms of wavelength.
This set of colors provides an information on the shape of the SED but effects of the SFH are also expected on wider scales in terms of wavelength.
As discussed in \cite{Ciesla18}, discrepancy between the UV and NIR emission assuming a smooth delayed-$\tau$ SFH is the signature that we are looking for indicating a possible change in the recent SFH.
To be able to probe these effects, we also normalize each photometric band to the Ks flux and add this set of colors to the previous one.
Finally, we set the flux ratios FUV/NUV and FUV/Ks to be $0$ when $z>0.68$ to account for the missing FUV flux density due to the Lyman break at these redshifts. 

\section{\label{stat}Statistical approach}
\newcommand{\xobs}{x_\text{obs}}
In this section, we present the statistical approach that we will use to infer the most suitable SFH from photometric data. 
This new approach is applied to the sample described in Sect.~\ref{sample} as a pilot study but can be applied to other datasets and to test other properties than the SFH.

\subsection{Statistical modeling}
\label{sub:stat_modeling}

As explained in the previous section, we want to distinguish between two SFH models: the first one is the smooth delayed-$\tau$ SFH, or SFH model $m=0$, and the second is the same with a flexibility in the last 500\,Myr, or SFH model $m=1$, as presented in Sect.~\ref{c18}.
The smooth delayed-$\tau$ SFH is thus a specific case of the flexible SFH obtained when there is no burst nor quenching ($r_{\rm{SFR}}=1$).

Let $\xobs$ denote the broadband data
collected about a given galaxy. The statistical issue of deciding which
SFH  better fits the data can be  seen as the Bayesian testing procedure distinguishing between both hypotheses
\[
  H_0:\ r_\text{SFR}=1 \quad \text{vs}
  \quad H_1:\ r_\text{SFR} \neq 1.
\]
The procedure will decide in favor of a possible change in the recent history when
$r_\text{SFR}$ is significantly different from $1$ based on the data $\xobs$.
Conducting a Bayesian testing procedure based on the
data $\xobs$ of a given galaxy is exactly the same thing as the Bayesian model choice distinguishing between two nested statistical models \citep{Robert07}.

The first statistical model ($m=0$), that is the delayed-$\tau$ SFH, is composed as follow: Let $\theta_0$  denote the vector of all parameters necessary to compute the mock SED, denoted $\text{SED}(\theta_0)$. In particular $\theta_0$ includes the parameters of the SFH. We denote $p(\theta_0|m=0)$ the prior distribution over the parameter space for this statistical model.
Likewise for the second SFH model : let $\theta_1$ =($\theta_0, r_\text{SFR},t_\text{flex})$ be the vector of all parameters for the delayed-$\tau$+flex SFH. This vector includes the same parameters as for the previous SFH, plus two added parameters $r_\text{SFR}$  and $t_\text{flex}$. Let $p(\theta_1|m=1)$ be the prior distribution over the parameter space for the second model. 
We furthermore add a prior probability on the SFH index, $p(m=1)$ and $p(m=0)$, which are both $0.5$ if we want to remain noninformative.

Finally, we assume a Gaussian noise. Thus, the likelihood $p(\xobs|\theta_m, m)$ of $\theta_m$ given $\xobs$ under the statistical model $m$ is a multivariate Gaussian distribution, centered on  $\text{SED}(\theta_m)$ with a diagonal covariance matrix. The standard deviations are set to $0.1\times\text{SED}(\theta_m)$ because of the assumed value of SNR in the observations. In particular, it means that, up to constant, the loglikelihood is the negative $\chi^2$-distance between the observed SED and the mock $\text{SED}(\theta_m)$:
\begin{align}
    p(\xobs|\theta_m, m) &\propto \exp\left(-\frac12 \chi^2\Big(\xobs, \text{SED}(\theta_m)\Big) \right), \quad\text{where } \notag
    \\ \label{eq:gaussian_noise}
    \chi^2\Big(\xobs, \text{SED}(\theta_m)\Big) &= \sum_{j=1}^J \frac{\Big(\xobs(\lambda_j) - \text{SED}(\theta_m, \lambda_j) \Big)^2}{0.1\text{SED}(\theta_m, \lambda_j)}.
\end{align}

\subsection{\label{Bmc}Bayesian model choice}
\label{sub:bayes}

Bayesian model choice \citep{Robert07} relies on the evaluation of the
posterior probabilities $p(m|\xobs)$ which, using Bayes formula, is given by
\begin{equation}
  p(m|\xobs) = \frac{p(m) p(\xobs|m)} {\displaystyle\sum_{m'} p(m') p(\xobs|m')},
\end{equation}
where
\begin{equation} \label{eq:evidence}
   p(\xobs|m)=\int
  p(\xobs|\theta_m,m)p(\theta_m|m)
  d\theta_m
\end{equation}
is the likelihood integrated over the prior distribution of the $m$-th statistical model. Seen as a function of $\xobs$, $p(\xobs|m)$ is called the evidence or the integrated likelihood of the  $m$-th model.

Bayesian model choice procedure innately embodies Occam's razor. This principle consists in choosing the simplest model as long as it is sufficient to explain the observation{\footnote{  Indeed,  the evidence $p(\xobs|m)$ is a normalized probability density, that represents the distribution of datasets drawn from the $m$-th model, whatever the value of the parameter $\theta_m$ from its prior distribution. If models $m=0$ and $m=1$ are nested, the region of the data space of non-negligible probability under model $m=0$ has also a non-negligible probability under model $m=1$. Moreover, since model $m=1$ can fit to much more datasets, the probability density $p(\xobs|m=1)$ is much more diffuse than the density $p(\xobs|m=0)$. Hence, we expect for datasets $x$ that can be explained by both models $m=0,1$ that
$
p(x|m=1) \le p(x|m=0).
$
If the prior probabilities $p(m=0)$ and $p(m=1)$ of both models are equal, it implies that, for datasets $\xobs$ that can be explained by both models,
$p(m=1|x) \le p(m=0|x).$
}}.
In this study, the two parametric SFHs are nested: when the parameter $r_\text{SFR}$ of an SFH  $m=1$ (flex $+$ delayed-$\tau$) is set to $1$, we have an SFH that is also in the model $m=0$ (delayed-$\tau$).
Because of Occam's razor, if we choose the SFH with highest posterior probability when analyzing an observed SED $\xobs$ that can be explained by both SFHs, we choose the simplest model $m=0$.

To analyse the dataset $\xobs$, it remains to compute the posterior probabilities. In our situation, the evidence of the statistical model $m$ is intractable. It  means that it cannot be easily evaluated numerically. Indeed, the function that computes $\text{SED}(\theta_m)$ given $m$ and  $\theta_m$  is fundamentally a black-box  numerical function.  

There are two methods to solve this problem. First, we can use a Laplace approximation of the integrated likelihood. The resulting procedure is the one that choose the SFH with the smallest Bayesian Information Criterion (BIC). Denoting $\hat{\theta}_m$ the maximum likelihood estimate under the SFH $m$, $\chi^2$ the non-reduced $\chi^2$-distance of the fit, $k_m$ the degree of freedom of model $m$, and $n$ the number of observed photometric bands, the BIC of SFH $m$ is given by
\begin{align} 
        \text{BIC}_m & = -2 \max_{\theta_m}\ln p(\xobs|\theta_m,m) + k_m \times \ln(n), 
        \notag
        \\
        & \label{BIC} = 
        \chi^2 \Big(\text{SED}(\hat{\theta}_m),\, x_\text{obs}\Big) + k_m \times \ln(n).
\end{align}
Choosing the model with the smallest BIC is therefore an approximate method to find the model with the highest posterior probability. The results of \citet{Ciesla18} based on BIC are justified on this ground. But the Laplace approximation assumes that the number of observed photometric bands $n$ is large enough. Moreover, determining the degree of freedom $k_m$ of a statistical model can be a complex question. For all these reasons, we expect to improve the method of \citet{Ciesla18} based on BIC in the present paper.

Clever Monte Carlo algorithms to compute the evidence, Equation \eqref{eq:evidence}, of  each statistical model will give us a much sharper approximation of the posterior probabilities of each SFH . We decided to rely
on Approximate Bayesian Computation \citep[ABC, see e.g. ][]{marin2012approximate,sisson2018handbook} to compute $p(m|\xobs)$.We could have considered other methods \citep{vehtari2012survey} such as 
bridge sampling, reversible jump MCMC, nested sampling, etc. But these methods require separate runs of the algorithm to analyze each galaxy, and probably more than a few minutes per galaxy. We expect to design a faster method here with ABC.

Finally,to interpret the results, we rely on the Bayes factor of the delayed-$\tau$+flex SFH ($m=1$)
against the delayed-$\tau$SFH ($m=0$)given by
\newcommand{\opn}{\operatorname}
\[
  \opn{BF}_{1/0}(\xobs) = \frac{p(\xobs|1)}{p(\xobs|0)} = \frac{p(1|\xobs)}{p(0|\xobs)} =
  \frac{p(1|\xobs)}{1- p(1|\xobs)}.
\]
The computed value of the Bayes factor is compared to standard thresholds established by Jeffreys \citep[see, e.g., ][]{Robert07} in order to evaluate the strength of the evidence in favor of delayed-$\tau$+flex SFH if $\opn{BF}_{1/0}(\xobs)\ge 1$. Depending on the value of the Bayes factor, Bayesian statisticians are used to say that the evidence in favor of model $m=1$ is either \textit{barely worth mentioning} (from $1$ to $\sqrt{10}$) or \textit{substantial} (from $\sqrt{10}$ to $10$) or \textit{strong} (from $10$ to $10^{3/2}$) or \textit{very strong} (from $10^{3/2}$ to $100$) or \textit{decisive} (larger than $100$).

\subsection{\label{ABC}The Approximate Bayesian Computation method}
To avoid the difficult computation of the evidence, Equation \eqref{eq:evidence}, of model $m$ and
get a direct approximation of $p(m|\xobs)$, we resort to the family of
methods named ABC model choice
\citep{marin2018likelihood-free}.

\begin{table}
\caption{\label{basic_abc} Basic ABC model choice algorithm that aims at computing the posterior probabilities of statistical models in competition to explain the data.
}
\rule{\columnwidth}{.2mm}
\textbf{Input}:
\begin{itemize}
    \item $\xobs$, the observed SED we want to analyse
    \item $p(m)$, prior probability of the $m$-th statistical model
    \item $p(\theta_m|m)$, prior distribution of parameter $\theta_m$ of the $m$-th statistical model
    \item $p(x|\theta_m, m)$, probability density  of a SED $x$ given the $m$-th statistical model, and the parameter $\theta_m$, see Eq.~\eqref{eq:gaussian_noise}
    \item $N$, number of simulations from the prior 
    \item $S(x)$, a function that computes the summary statistics of a SED $x$
\end{itemize}
\textbf{Output}:\\
An approximation $\hat p(m|\xobs)$ of the posterior probability of the $m$-th statistical model given the observed data for all $m$.
\begin{enumerate}[1\ ] 
    \item For $i = 1$ to $N$
    \item \quad Generate $m^i$ from the prior $p(m)$
    \item \quad Generate $\theta_m^i$ from the prior $p(\theta_m|m)$
    \item \quad Generate $x^i$ from the model $p(x|\theta_m, m)$
    \item \quad Compute $S(x^i)$ and store $(m^i, \theta_m^i, S(x^i))$
    \item End For
    \item Compute $\hat p(m|\xobs)$ with Eq.~\eqref{eq:phat} for all $m$
\end{enumerate}
\rule{\columnwidth}{.2mm}
\end{table}

The main idea behind the ABC framework is that we can avoid the evaluation of the likelihood and directly estimate a posterior probability by relying on $N$
random simulations $(m^i, \theta_m^i, x^i)$, $i=1,\ldots,N$ from the joint distribution $p(m)p(\theta_m|m) p(x|\theta_m, m)$. Here simulated $(m^i, \theta_m^i, x^i)$ are obtained as follow: first, we draw a SFH $m^i$ at random, with the prior probability $p(m^i)$; then we draw $\theta_m^i$ according to the prior $p(\theta_m^i|m^i)$; finally we compute the mock $\text{SED}(\theta_m^i)$ with CIGALE and add a Gaussian noise to the mock SED to get $x^i$. This last step is equivalent to sampling from $p(x^i|\theta_m^i, m^i)$ given in \eqref{eq:gaussian_noise}. Basically, the posterior distribution $p(m|\xobs)$ can be approximated by the frequency of SFH $m$ among the simulations close enough to $\xobs$. 

To measure how close $x$ is from $\xobs$, we introduce the distance between vectors of summary statistics $d\big(S(x), S(\xobs)\big)$ and we set a threshold $\epsilon$: simulations $(m,\theta_m, x)$ that satisfy $d\big(S(x), S(\xobs)\big)\le \epsilon$ are considered ``close enough" to $\xobs$. The summary statistics $S(x)$ are primarily introduced as a way to handle feature extraction, whether it is for dimensionality reduction or for data normalization. For the present study, the components of the vector $S(x)$ are flux ratios from the SED $x$, chosen for normalization purposes.
Mathematically speaking, $p(m=1|\xobs)$ ies thus approximated by
\begin{equation} \label{eq:phat}
\hat p(m|\xobs) = \frac{\displaystyle
\sum_{i=1}^N \mathbf 1\{m^i = m\} \mathbf 1\Big\{d\big(S(x^i), S(\xobs)\big)\le \epsilon\Big\}
}{\displaystyle
\sum_{i=1}^N \mathbf 1\Big\{d\big(S(x^i), S(\xobs)\big)\le \epsilon\Big\}
}.
\end{equation}
The resulting algorithm, named basic ABC model choice, is given in Table~\ref{basic_abc}. Finally, note that, if $k$ is the number of simulations close enough to $\xobs$,  the last step of Table~\ref{basic_abc} can be seen as a $k$-nearest neighbor ($k$-nn) method predicting $m$ based on the features (or covariates) $S(x)$. 

The $k$-nn can be replaced by other machine learning algorithms to obtain sharper results. Indeed, the $k$-nn is known to perform poorly when the dimension of $S(x)$ is larger than $4$. For instance, \citet{pudlo2014} decided to rely on the method called Random Forest \citep{breiman2001random}. The machine learning based ABC algorithm is given in Table~\ref{ml_abc}. All machine learning models given below are classification methods. In our context, they aim at separating the simulated datasets $x$ depending on the SFH ($m=0\text{ or } 1$) that was used to generate them. The machine learning  model is fitted on the catalog of simulations $(m^i, \theta_m^i, x^i)$, that is to say, it learns how to predict  $m$ based on the value of $x$. To this purpose, we fit a function $\hat p(m=1|x)$ and perform the classification task on a new dataset $x'$ by comparing the fitted $\hat p(m=1|x')$ to $1/2$: if   $\hat p(m=1|x')>1/2$, the dataset $x'$ is classified as generated by SFH $m=1$; otherwise, it is classified as generated by SFH $m=0$. The function $\hat p(m=1|x')$  depends on some internal parameters not explicitly shown in the notation. For example, this function can be computed with the help of a neural network. A neuron here is a mathematical function that receives inputs and produces an output based on a weighted combination of the inputs; each neuron processes the received data and transmits its output downstream in the network. Generally, the internal parameters $(\phi, \psi)$ are of two kinds: the coordinates of $\phi$ are optimized on data with a specific algorithm, and the coordinates of $\psi$ are called tuning parameters (or hyperparameters). For instance, with neural networks $\psi$ represents the architecture of the network and the amount of dropout; $\phi$ represents the collection of the weights in the network.

The gold standard machine learning practice is to split the catalog of data into three parts: the training catalog and the validation catalog, that are both used to fit the machine learning models, and the test catalog that is used to compare the algorithms fairly and get a measure of the error committed by the models. Actually each fit requires two catalogs (training and validation) because modern machine learning models are fitted to the data with a two step procedure. We detail the procedure for a simple dense neural network and refer to Appendix~\ref{apFit} for the general case.
The hyperparameters we consider are the number of hidden layers, the number of nodes in each layers, and the amount of dropout. We fix a range of possible values for each hyperparameters (see table \ref{tab:calibration}). We select a possible combination of hyperparameters  $\psi$, and train the obtained neural network on the training catalog. Once the weights $\phi$ are optimized on the training catalog, we evaluate the given neural network on the validation catalog and associate the obtained classification error with the combination of hyperparameters used. We follow the same training and evaluating procedure for several hyperparameters combinations  $\psi$, and we select the one obtaining the lowest classification error.
At the end of the process, we evaluate the classification error on the test catalog using the selected combination of hyperparameters $\hat\psi$

The test catalog is willingly left out during the training and the tuning of the machine learning methods. Indeed, the comparison of the accuracy of the approximation returned by each machine learning method on the test catalog ensures a fair comparison between the methods, on data unseen during the fit of $\hat p_{\hat \psi}(m|x)$. 

In this pilot study, we tried different machine learning methods and compared their accuracy:\begin{itemize}
    \item logistic regression and linear discriminant analysis \citep{friedman2001elements}, that are almost equivalent linear models, and serve only as baseline methods,
    \item neural networks with $1$ or $3$ hidden layers, the core of deep learning methods, that have proved to get sharp results on various signal datasets (images, sounds)
    \item classification tree boosting \citep[with XGBoost, see ][]{chen2016xgboost}, which is considered as state-of-the-art methods in many applied situations, and is often the most accurate algorithm when correctly calibrated on a large catalog.
\end{itemize}
We did not try Random Forest since it cannot be run on a simulation catalog of size as large as the one we are relying on in this pilot study ($N = 4 \times 10^6$).
Indeed the motivation of the proposed methodology is to bypass the heavy computational burden of MCMC based algorithms to perform statistical model choice. In this study, Random Forest was not able to fulfill this aim unlike the classification methods given above.

\begin{table}
\caption{\label{ml_abc}Machine learning based  ABC model choice algorithm that aims at computing the posterior probability of two statistical models in competition to explain the data}
\rule{\columnwidth}{.2mm}
\textbf{Input and output}: same as Table~\ref{basic_abc}
\begin{enumerate}[\bf 1\ ] \normalsize
    \item Generate  $N$ simulations $(m^i,\theta_m^i, x^i)$  from the joint distribution {$p(m) p(\theta_m|m)p(x| \theta_m,m)$}
    \item Summarize all simulated datasets (photometric SED) $x^i$ with $S(x^i)$  and store all simulated  $(m^i, \theta_m^i, S(x^i))$  into a large catalog
    \item Split the catalog into three parts: training, validation and test catalogs
    \item Fit each machine learning method on the training and validation catalogs to approximate $p(m= 1|S(x))$ with $\hat p_{\hat \psi}(m=1|x)$
    \item Choose the best machine learning method by comparing their classification errors on the test catalog
    \item Return the approximation $\hat p(m=1|x_\text{obs})$ computed with the best method
\end{enumerate}
\rule{\columnwidth}{.2mm}

\end{table}

\subsection{ \label{prior} Building synthetic photometric data }

To compute or fit galaxies' SEDs with CIGALE, one has to provide a list of prior values for each model's parameters. 
The comprehensive module selection in CIGALE allows to specify entirely the SFH and how the mock SED is computed. 
The list of prior values for each module's parameters specifies the prior distribution $p(\theta_m|m)$. 
CIGALE uses this list of values or ranges to sample from the prior distribution by picking values on $\theta_m$ on a regular grid. 
This has the inconvenient of: being very sensitive to the number of parameters (if $d$ is the number of parameters, and if we assume $10$ different values for each parameter, the size of the grid is $10^d$); producing simulations that are generated with some parameters that are equals.
Instead, in this study, we advocate in favor of drawing values of all parameters at random from the prior distribution, which is uniform over the specified ranges or list of values.
The ranges for each model parameters are chosen to be consistent with those used by \cite{Ciesla18}. 
In particular, the catalog of simulations drawn at line~1 in Table~\ref{ml_abc} follow this rule. 
Each SFH  (the simple delayed-$\tau$ or the delayed-$\tau$ $+$ flex) is then convolved with the stellar population models of \cite{BruzualCharlot03}.
The attenuation law described in \cite{CharlotFall00} is then applied to the SED.
Finally CIGALE convolves each mock SED  into a COSMOS-like set of filters described in Table~\ref{bands}.

\begin{table}
   \centering
   \caption{Prior range of the parameters used to  generate the simulation table of  SEDs with redshift between 0.5 and 1.}
   \begin{tabular}{l c }
   \hline\hline
   \textbf{Parameter} & \textbf{Value} \\
   \hline
   \multicolumn{2}{c}{\textbf{Delayed-$\tau$ SFH}}\\[1mm]  
   $age$ (Gyr) & $\big[$0.5;\ 9$\big]$        \\
   $\tau_{main}$  (Gyr) & $\big[$0.1;\ 10$\big]$ \\[1mm]
   \multicolumn{2}{c}{\textbf{Flexible delayed-$\tau$ SFH}}\\[1mm]  
   $age$ (Gyr) &  $\big[$0.5;\ 9$\big]$       \\
   $\tau_{main}$  (Gyr) & $\big[$0.1;\ 10$\big]$\\
   $age_{flex}$ (Myr) & 10, 100, 450\\
   $\log r_{\rm{SFR}}$  & $\big[$-6;\ 6$\big]$\\[1mm]
   \multicolumn{2}{c}{\textbf{Dust attenuation}}\\[1mm]  
   $A_V$      &  $\big[$0.1;\ 4$\big]$  \\
   \hline
   \label{mockparam}
   \end{tabular}
\end{table}

\section{Application to synthetic photometric data} \label{synthetic}

We first applied our methodology on simulated photometric data to evaluate its accuracy. The main interest of such synthetic data is that we control all parameters (flux densities, colors, physical parameters). The whole catalog of simulations was composed of $4\times 10^6$ simulated datasets. We split this catalog at random into three parts, as explained in Sect.~\ref{ABC}, and add an extra catalog for comparison with CIGALE:
\begin{itemize}
\item $3.6\times 10^6$ sources $(90\%)$ to compose the training catalog,
\item $200,000$ sources $(5\%)$ to compose the validation catalog,
\item $200,000$ sources $(5\%)$ to compose the test catalog,
\item $30,000$ additional sources to compose the extra catalog for comparison with CIGALE.
\end{itemize}
The size of the extra catalog is much smaller to limit the amount of computation time required by CIGALE to run its own algorithm of SED fitting.

\subsection{\label{train}  Calibration and evaluation of the machine learning methods on the simulated catalogs} 

\begin{table*}
\caption{Calibration and test of machine learning methods}           
\label{tab:calibration}      
\centering          
\begin{tabular}{r c c r r}     
\hline\hline       
\textbf{Method} & \textbf{Tuning parameter} & \textbf{Explored range} & \textbf{Best value} & \textbf{Error rate (\%)}\\ 
\hline
Logistic regression & $\emptyset$ & & & $30.27$ \\[2mm]
Linear Discriminant Analysis & $\emptyset$ & & & $30.43$ \\[2mm]
$k$-nearest neighbors & $k$ & $[3600,\  180000]$ & $5000$ & $23.79$\\[2mm]
$1$-layer neural network & dropout & $[0.1,\ 0.5 ]$ & 0.2 & $22.51$\\
   & nodes in each layer & $[16,\ 256]$ & $128$ & ---\\[2mm]
$3$-layer neural network & dropout & $[0.1,\ 0.5 ]$ & 0.2 & $21.06$\\
   & nodes in each layer & $[16,\ 256]$ & $128$ & ---\\[2mm]
Tree boosting (XGBoost) & number of trees (nround) & $[100,\ 1000]$ & 400 & $20.98$\\
   & depth of each tree (max\_depth) &  $[4,\ 15]$ & 12 & ---\\
   & learning rate (eta) & $[0.01,\ 0.2]$ & $0.1$ & ---\\
\hline                  
\end{tabular}\\
The best value of each tuning parameter was found by comparing error rates on the validation catalog.\\ 
The error rate given in the last column is computed on the test catalog.
\end{table*}

In this section, we present the calibration of the machine learning techniques and their error rates on the test catalog. We then try to interpret the results given by our methodology.

As described in Sect.~\ref{ABC}, we trained and calibrated the machine learning methods on the training and validation catalog. The results are given in Table~\ref{tab:calibration}. Neither Logistic regression nor Linear Discriminant Analysis have tuning parameters that need to be calibrated on the validation catalog. The error rate of these techniques are about $30\%$ on the test catalog. But the modern machine learning methods ($k$-nearest neighbors, neural networks and tree boosting) have been calibrated on the validation catalog. The best value of the explored range for $\psi$ were found by comparing error rates on the validation catalog and are given in Table~\ref{tab:calibration}. The error rates of these methods on the test catalog vary between $24\%$ and $20\%$. 
Thus, it is clear that there is a significant gain to use non-linear methods. But we see no obvious use in training a more complex algorithm (such as a deeper neural network) for this problem, although it could become useful when increasing the number of photometric bands and the redshift range.
Finally, we favor XGBoost for our study. Indeed, while neural networks could probably be tuned more precisely to match or exceed its performances, we find XGBoost easier to tune and to interpret.

Machine learning techniques that fit $\hat p_{\hat \psi}(m|x)$ are often affected by some bias and may require some correction \citep{NiculescuMizil12}.Such classification algorithms compare the estimated probabilities of $m$ given $x$ and return the most likely $m$ given $x$. The output $m$ can be correct even if the probabilities are biased towards $0$ for small probabilities or towards $1$ for large probabilities. A standard reliability check shows no such problem for our XGBoost classifier. To this aim, the test catalog is divided into $10$ bins: the first bin is composed of simulations with a predicted probability $ \hat p(m=1 | \xobs)$ between $0$ and $0.1$, the second with $\hat p(m=1 | \xobs)$ between $0.1$ and $0.2$\ldots The reliability check procedure ensures that the frequency of the SFH $m=1$ among the $k$-th bin falls within the range $[(k-1)/10;k/10]$, because the $\hat p(m=1 | \xobs)$ predicted by XGBoost are between $(k-1)/10$ and $k/10$. 

\begin{figure*} 
  	\includegraphics[width=\textwidth]{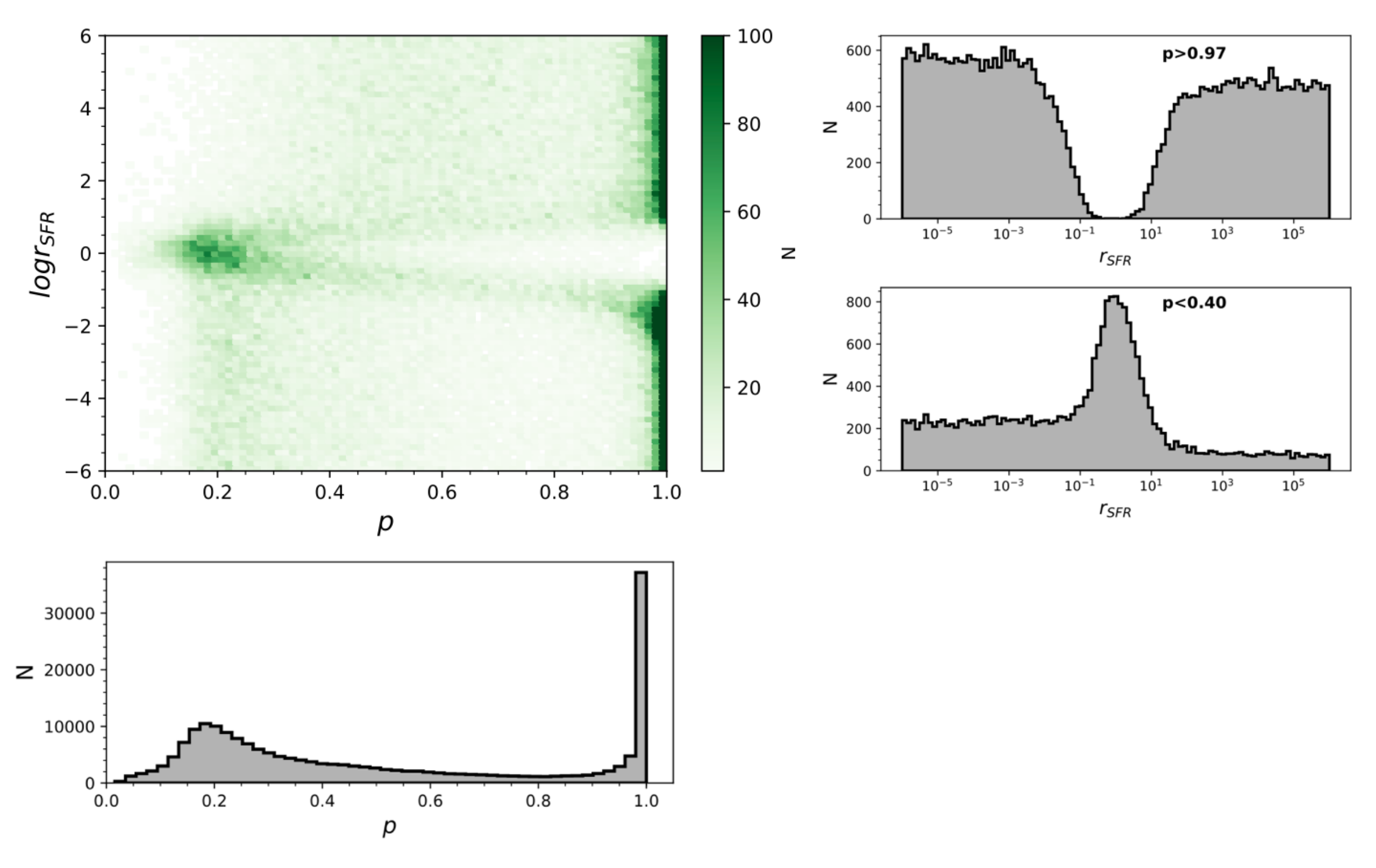}
  	\caption{\label{pmock} Study of the statistical power of  $\hat p(m=1 | \xobs)$ to detect short-term variations with respect to the value of $r_{\mathrm{SFR}}$. \textbf{Top left panel:} Joint distribution of $\hat p(m=1 | \xobs)$ and $r_{\mathrm{SFR}}$. \textbf{Bottom left panel:}  Distribution of $\hat p(m=1 | \xobs)$ obtained with $x$ coming from the test catalog. \textbf{Right panels:} Marginal distributions of $r_{\mathrm{SFR}}$ for mock sources with $\hat p(m=1 | \xobs)>0.97$  (top right panel) and for mock sources with $\hat p(m=1 | \xobs)<0.4$ (bottom right panel).}
\end{figure*}

We studied the ability of our methodology to distinguish the SFH of the simulated sources of the test catalog.
The top panel of Fig.~\ref{pmock} shows the distribution of $\hat p(m=1 | \xobs)$ when $x$ varies in the test catalog. 
Naively, a perfect result would have half of the sample with $p=1$ and the other half with $p=0$.
In fact, when $m=0$, the SFH $m=1$ is also suitable since the models are nested. In this case, Occam's razor favors the model $m=0$, and $\hat p(m=1 | \xobs)$ must be less than $0.5$, see Sect.~\ref{sub:bayes}. On the contrary, for the SEDs solely explained by the SFH model $m=1$, $\hat p(m=1 | \xobs)$ is close to $1$. 

The distribution (Fig.~\ref{pmock}, bottom left panel) has two peaks, one centered around $p=0.2$ and one between $0.97$ and $1$.
This peak at $0.2$, and not $0$, is expected when one of the model proposed to the choice is included in the second model.
In the distribution of the $\hat p(m=1 | \xobs)$, 20$\%$ of the sources have a value higher than 0.97 and 52$\%$ lower than 0.4. 
In the right panels of Fig.~\ref{pmock}, we show the distribution of {$r_{\text{SFR}}$ } for the galaxies $x$ with $\hat p(m=1 | \xobs)>0.97$.
With a perfect method, galaxies with $r_{\mathrm{SFR}}\neq1$ should have $\hat p(m=1 | \xobs)=1$.
Here we see indeed a deficit of galaxies around {$p=1$}, however the range of affected $r_{\text{SFR}}$ goes from $0.1$ to $10$. 
Therefore, the method is not able to identify galaxies having a variability of its SFR if this variability is only $0.1$ to $10$ times the SFR before the variability began.
In other words, the method is sensitive to $|\log r_{\mathrm{SFR}}|>1$.
This is confirmed by the distribution of $r_{\mathrm{SFR}}$ for galaxies with $p<0.40$ (Fig.~\ref{pmock}, bottom panel).
However, there are sources with a $|\log r_{\mathrm{SFR}}|>1$ associated to low values of $\hat p(m=1 | \xobs)$.
The complete distribution of $r_{\mathrm{SFR}}$ as a function of $\hat p(m=1 | \xobs)$  is shown in Fig.\ref{pmock}.

\subsection{\label{importance}Importance of particular flux ratios}
    
We try to find which part of the dataset $x$ influences the most on the choice of SFH given by our method. The analyse of $x$ relies entirely on the summary statistics $S(x)$, the flux ratios. Hence, we tried to understand which flux ratios are most discriminant for the model choice. We wanted to check that the method is not based on a bias of our simulations and wanted to assess which part of the data could be removed without losing crucial information.

We use different usual metrics \citep[e.g. ][]{friedman2001elements,chen2016xgboost} to assess the importance of each flux ratio in the machine learning estimation of $\hat p(m=1|x)$.
Those metrics are used as indicators of the relevance of each flux ratio for the classification task. 
As expected, the most important flux ratios for our problem involve the bands at shortest wavelength (FUV at $z<0.68$ and NUV above, as FUV is no longer available), normalized by either Ks or u.
This is expected as these bands are known to be sensitive to SFH \citep[e.g.,][]{Arnouts13}.
We see no particular pattern in the estimated importance of the other flux ratios. 
They are all used for the classification and removing any of them decreases classification accuracy, except for IRAC1/Ks whose importance is consistently negligible across every considered metric. 

We also test if the UVJ selection used to classify galaxies according to their star formation activity \citep[e.g.,][]{Wuyts07,Williams09} is able to probe the kind of rapid and recent SFH variations we are investigating in this study. We train an XGBoost classification model using only u/V and V/J in order to evaluate the benefits of using all available flux ratios. This results in a severe increase in classification error, going from $21.0\%$ using every flux ratios to $35.8\%$. 

\subsection{\label{mock}Comparison with SED fitting methods based on BIC}

In this section, we compare the results obtained with the ABC method to those obtained with a standard SED modeling.
The goal of this test is to understand and quantify the improvement that the ABC method brings in terms of accuracy of the results.
We use the simulated catalog of 30,000 sources, described in the beginning of this section,  for which we control all parameters.

The ABC method is also used on this extra catalog.
This test is very similar to the training procedure described in Sect.~\ref{train}.
Indeed, with this extra catalog, the ABC method has an error  rate of 21.2\% compared to 21.0\% with the previous test sample. 

\begin{table}
   \centering
   \caption{Input parameters used in the SED fitting procedures with CIGALE.}
   \begin{tabular}{l  l }
   \hline\hline
   Parameter & Value \\ 
   \hline\hline
   \multicolumn{2}{c}{delayed-$\tau$ SFH}\\  
   \hline
   $age$ (Gyr) & $[0.5;9]$, 15 values linearly sampled         \\
   $\tau_{main}$  (Gyr) & $[0.1;10]$, 15 values linearly sampled\\
   \hline\hline
   \multicolumn{2}{c}{Flexible delayed-$\tau$ SFH}\\  
   \hline
   $age$ (Gyr) &  $[0.5;9]$, 15 values linearly sampled        \\
   $\tau_{main}$  (Gyr) & $[0.1;10]$, 15 values linearly sampled\\
   $age_{flex}$ (Myr) & 10, 100, 450\\
   $\log r_{\rm{SFR}}$  & $[-6;6]$, 12 values linearly sampled     \\ 
   \hline\hline
   \multicolumn{2}{c}{Dust attenuation}\\  
   \hline
   $A_V^{ISM}$      &  $[0.1;4]$, 10 values linearly sampled\\
   \hline
   \hline
   \label{input_param}
   \end{tabular}
\end{table}

CIGALE is run on the test catalog as well.
The set of modules is the same as those used to create the mock SEDs, however the parameters used to fit the test catalog do not include the input parameters that were randomly chosen.
This test is intentionally thought to be simple and represent an ideal case scenario.
The error rate that will be obtained with CIGALE will therefore represent the best result achievable.

To decide whether a flexible SFH was preferable to a normal delayed-$\tau$ SFH using CIGALE, we adopt on whether a flexible SFH is preferred to a normal delayed-$\tau$ SFH, we adopt the method of \cite{Ciesla18} described in Sect.~\ref{c18}.
The quality of fit using each SFH is tested through the use of the Bayesian Information Criterion (BIC). 

In detail, the method that we use is the following:
First, we make a run with CIGALE using a simple delayed-$\tau$ SFH which parameters are presented in Table~\ref{input_param}.
A second run is then performed with the flexible SFH.
We compare the results and quality of the fits using one SFH or the other.
The two models have different number of degrees of freedom.
To take this into account, we compute the BIC presented in Sect.~\ref{Bmc} for each SFH.

We then calculate the difference between BIC$_{delayed}$ and BIC$_{flex}$ ($\Delta$BIC) and use the threshold defined by Jeffreys (Sect.~\ref{Bmc}) valid either for the BF and the BIC 
and also used in \cite{Ciesla18}: 
a $\Delta$BIC larger than 10 is interpreted as a strong difference between the two fits \citep{KassRaftery95}, with the flexible SFH providing a better fit of the data than the delayed-$\tau$ SFH.

\begin{figure}[!h] 
  	\includegraphics[width=\columnwidth]{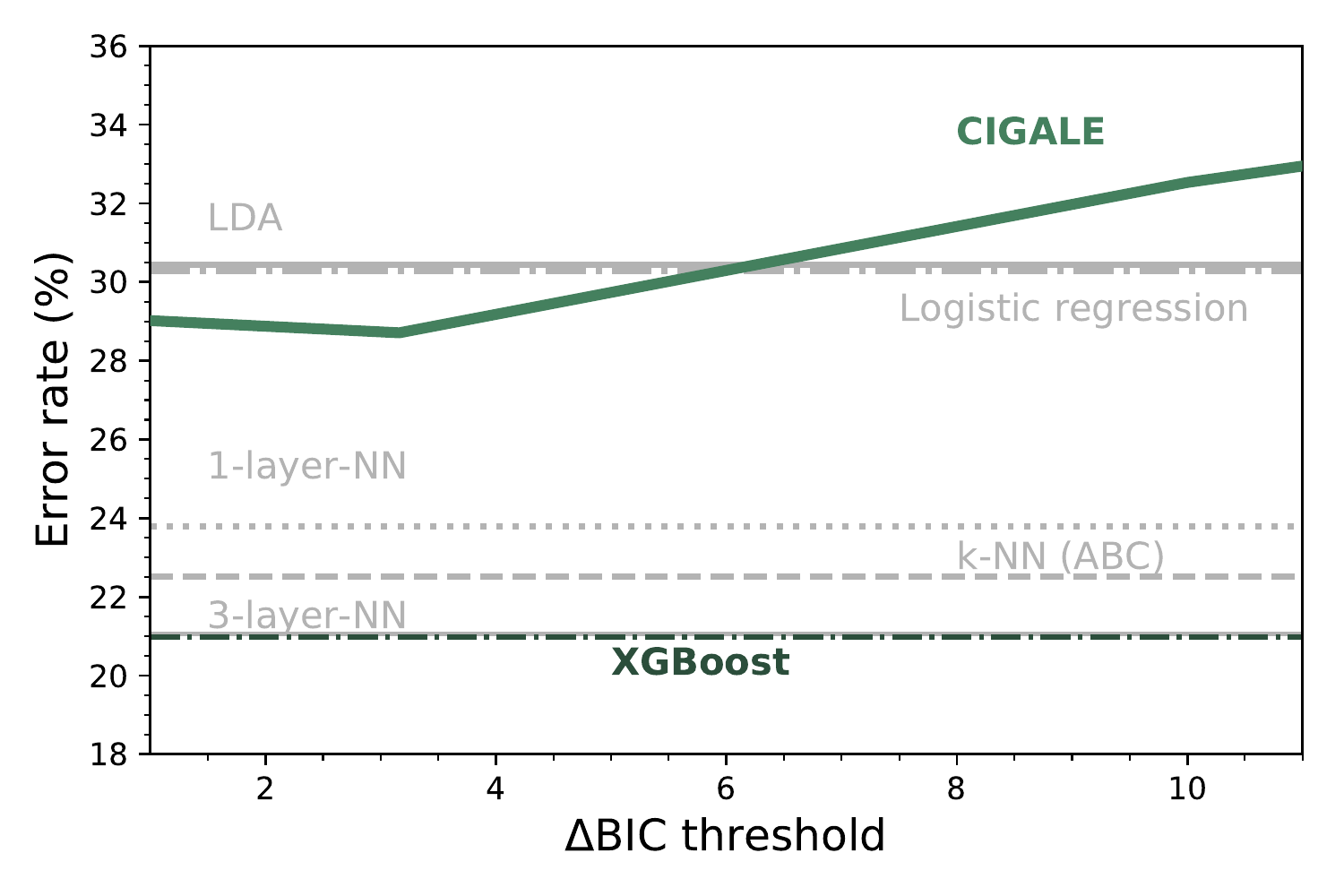}
  	\caption{\label{deltabicthresh}Error rate obtained with CIGALE as a funtion of $\Delta$BIC chosen threshold. For comparison we show the error rates obtained by the classification methods tested in Sect.~\ref{stat}.}
\end{figure}

We apply this method to the sample containing 15k sources modeled with a delayed-$\tau$ SFH and 15k modeled using a delayed-$\tau$+flexibility.
With this criteria, we find that the error rate of CIGALE, in terms of identifying SEDs built with a delayed-$\tau$+flex SFH, is 32.5\%.
This rate depends on the $\Delta$BIC threshold chosen and increases with the value of the threshold as shown in Fig.~\ref{deltabicthresh}.
The best value, 28.7\%, is lower than the error rate obtained from a logistic regression or a LDA (see Table~\ref{tab:calibration}) but significantly higher than the error rate obtained from our procedure using XGBoost (21.0\%)
In this best case scenario test for CIGALE, a difference of 7.7\% is substantial and implies that the ABC method tested in this study provides better results than a more traditional one using SED fitting.
When considering sources with $\Delta$BIC$>$10, i.e. sources for which the method using CIGALE estimates that there is a strong evidence for the flexible SFH, 95.4$\%$ are indeed SEDs simulated with the flexible SFH.
Using our procedure with XGBoost, and the Bayes factor corresponding threshold of 150 \citep{KassRaftery95}, we find that 99.7$\%$ of the sources' SFH are correctly identified.
The ABC method provides a cleaner sample than the CIGALE $\Delta$BIC based method.

\section{\label{real}Application on COSMOS data}

\begin{figure}[!h] 
  	\includegraphics[width=\columnwidth]{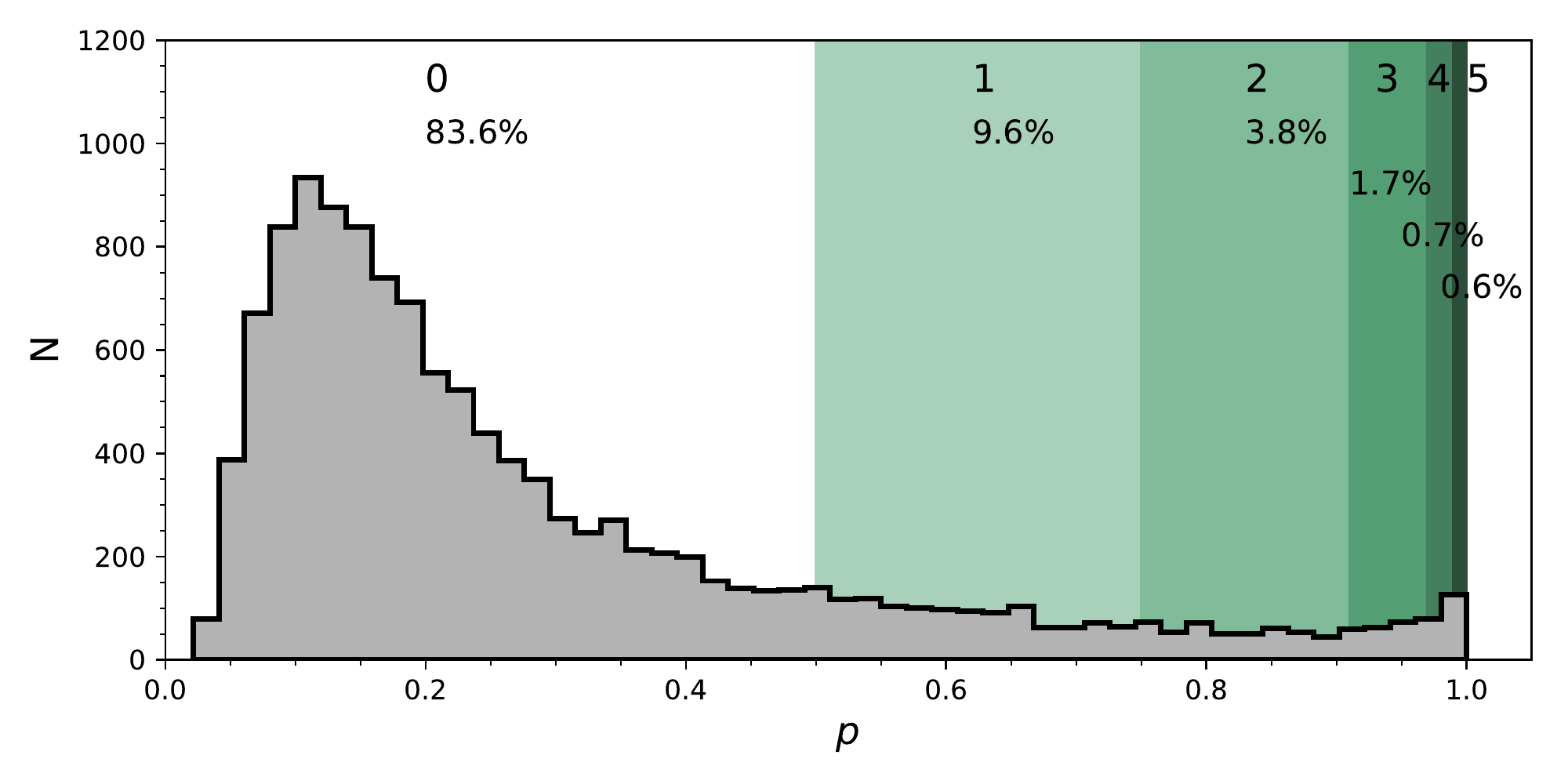}
  	\caption{Distribution of the predictions $\hat p(m=1 | \xobs)$ produced by our algorithm on the selected COSMOS data. Sources with a $\hat p(m=1 | \xobs)$ close to 1 tend to prefer the delayed-$\tau$+flex SFH while sources with lower $\hat p(m=1 | \xobs)$  favors a simple delayed-$\tau$ SFH. The green regions numbered from 1 to 5 indicate the Jeffreys scale of the Bayes factor, 1: Barely worth mentioning, 2: Substantial, 3: Strong, 4: Very strong, and 5: Decisive (detailed at the end of Sect. \ref{sub:bayes}). The percentage of sources in each grade is provided on the Figure and in Table~\ref{jeffreys}.}
  	\label{pred_cosmos}
\end{figure}

We now apply our method to the sample of galaxies drawn from the COSMOS catalog, which selection is described in Sect.~\ref{sample}.
As a result, we show the $\hat p(m=1 | \xobs)$ distribution obtained for this sample of observed galaxies in Fig.~\ref{pred_cosmos}.
We remind that the 0 value indicates that the delayed-$\tau$ SFH is preferred whereas $\hat p=1$ indicates that the flexible SFH is more adapted to fit the SED of the galaxy.
As a guide, we indicate the different grades of the Jeffreys scale and provide the number of sources in each grade in Table~\ref{jeffreys}.
The flexible SFH better models the observations of 16.4\% of our sample than the delayed-$\tau$ SFH.
However, it also means that for most of the dataset (83.6\%), there is no strong evidence for the necessity to increase the complexity of the SFH, a delayed-$\tau$ is sufficient to model the SED of these sources.

\begin{table}
   \centering
   \caption{Jeffreys scale and statistics of our sample.}
   \begin{tabular}{l l c c }
   \hline\hline
   Grade & Evidence against delayed-$\tau$ SFH & Number & \% \\ 
   \hline
   1    & Barely worth mentioning    & 1,187  & 9.6  \\
   2    & Substantial  & 466  &  3.8    \\
   3    & Strong  & 209  &   1.7   \\
   4    & Very strong     & 90   & 0.7 \\
   5    & Decisive     & 77  & 0.6  \\
   \hline
   \label{jeffreys}
   \end{tabular}
\end{table}

\begin{figure}[!h] 
  	\includegraphics[width=\columnwidth]{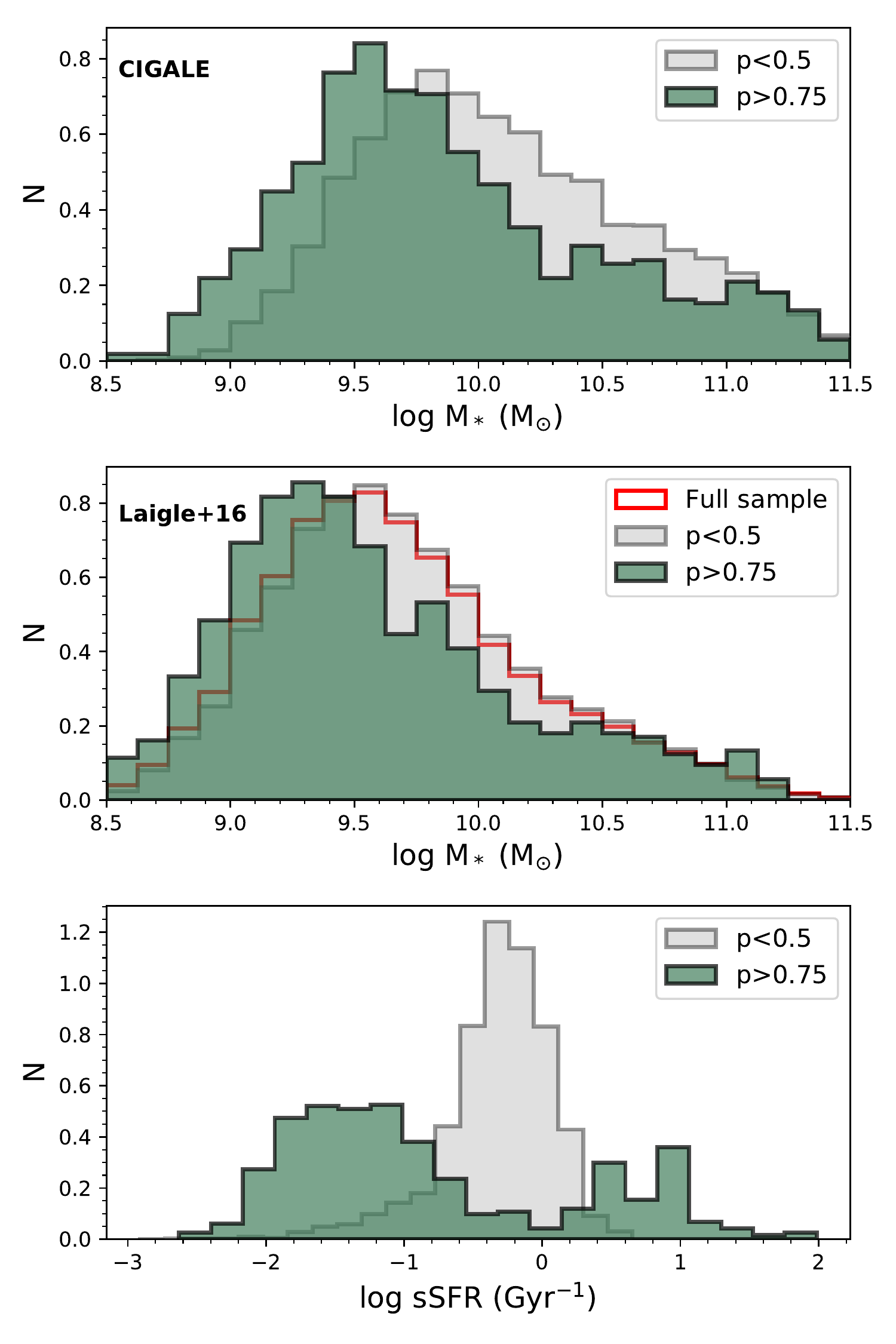}
  	\caption{\label{mstar}\textbf{Top panel:} Comparison of stellar mass distribution, obtained with CIGALE, for the sample of galaxies with $\hat p(m=1 | \xobs)>=0.75$  (green) and galaxies with $\hat p(m=1 | \xobs)<0.5$ (grey). \textbf{Middle panel:} Comparison of stellar mass distribution, obtained by \cite{Laigle16}, for the sample of galaxies with $\hat p(m=1 | \xobs)>=0.75$  (green) and galaxies with $\hat p(m=1 | \xobs)<0.5$  (grey). \textbf{Bottom panel:} Comparison of sSFR distribution for the sample of galaxies with $\hat p(m=1 | \xobs)>=0.75$  (green) and galaxies with $\hat p(m=1 | \xobs)<0.5$  (grey).}
  	\label{masses}
\end{figure}

To investigate the possible differences in terms of physical properties of galaxies according to their Jeffreys grade, we divide the sample of galaxies in two groups.
The first group corresponds to galaxies with$\hat p(m=1 | \xobs)<$ 0.5, galaxies for which there is no evidence for the need of a recent burst or quenching in the SFH, a delayed-$\tau$ SFH is sufficient to model the SED of these sources.
We select the galaxies of the second group imposing $\hat p(m=1 | \xobs)>$ 0.75, i.e. Jeffreys scale grades of 3, 4, or 5: from strong to decisive evidence against the normal delayed-$\tau$. 
In Fig.~\ref{mstar} (top panel), we show the stellar mass distribution of both subsamples.
Although the stellar masses obtained with either the smooth delayed-$\tau$ or the flexible SFH are consistent with each other, for each galaxies we use the most suitable stellar mass: if the galaxy has $\hat p(m=1 | \xobs)<0.5$ the stellar mass obtained from the delayed-$\tau$ SFH is used, and if the galaxy has$\hat p(m=1 | \xobs)>0.75$  the stellar mass obtained with the flexible SFH is used.
The stellar mass distribution of galaxies with a delayed-$\tau$ SFH is similar to the distribution of the whole sample, as shown in the middle panel of Fig.~\ref{masses}.
However, the stellar mass distribution of galaxies needing a flexibility in their recent SFH  shows a deficit of galaxies with stellar masses between 10$^{9.5}$ and 10$^{10.5}$\,M$_{\odot}$ compared to the distribution of the fool sample.
We note that at masses larger than 10$^{10.5}$\,M$_{\odot}$ the distribution are identical, despite a small peak at 10$^{11.1}$\,M$_{\odot}$.
To check if this results is not due to our SED modeling procedure and the assumptions we adopted, we show in the middle panel of Fig.~\ref{mstar} the same stellar mass distributions using this time the values published by \cite{Laigle16}.
The two stellar mass distributions, with the one of galaxies with $\hat p(m=1 | \xobs)>$ 0.75 peaking at a lower mass, are recovered.
This implies that these differences between the distributions are independent from the SED fitting method employed to determine the stellar mass of the galaxies.
We note that when the algorithm has been trained, only ratios of fluxes were provided to remove the normalization factor out of the method and the mock SEDs from which the flux ratios were computed were all normalized to 1\,M$_{\odot}$. 
The stellar mass is at first order a normalization through, for instance, the L$_K$-M$_*$ relation \citep[e.g.,][]{Gavazzi96}.
Using flux ratios, the algorithm had no information linked to the stellar mass of the mock galaxies. 
Nevertheless, applied to real galaxies the result of our procedure yields two different stellar mass distributions between galaxies identified as having smooth SFH and galaxies undergoing a more drastic episode (star formation burst or quenching).

In the bottom panel of Fig.~\ref{mstar}, we show the distribution in specific star formation rate (sSFR, $\mathrm{sSFR} \equiv \mathrm{SFR}/\mathrm{M_*}$) for the same two samples.
The distribution of galaxies with $\hat p(m=1 | \xobs)<0.5$  is narrow ($\sigma=0.39$ ) and has one peak at $\log \text{sSFR}=-0.32$ ($\text{Gyr}^{-1}$ ), clearly showing the MS of star forming galaxies.
Galaxies with high probability to have a recent strong variation of their SFH form a double-peaked distribution with one peak above the MS formed by galaxies with $\hat p(m=1 | \xobs)>$ 0.75 ($\log \mathrm{sSFR}=0.66$ ), corresponding to galaxies having experienced a recent burst, and a second peak at lower sSFRs than the MS, corresponding to sources having undergone a recent decrease of their star formation activity ($\log \mathrm{sSFR}=-1.38$ ).
In the sample of galaxies with $\hat p(m=1 | \xobs)>0.75$ , 28$\%$ of these sources are in the peak of galaxies experiencing a burst of star formation activity and 72$\%$ seem to undergo a rapid and drastic decrease of their SFR.
One possibility to explain this assymetry could be a bias produced by the algorithm, as shown in Fig.~\ref{pmock}, more sources with $\hat p(m=1 | \xobs)>$ 0.97 tend to be associated to low values of $r_{\mathrm{SFR}}$ than to $r_{\mathrm{SFR}}>1$.
However, in the case of the extra catalog, this disparity is 47$\%$ and 53$\%$ for high and low $r_{\mathrm{SFR}}$, respectively.

The distribution of the two samples in terms of sSFR indicates that, to be able to reach the sSFR of galaxies that are outside the MS, one had to take into account a flexibility in the SFH of galaxies when performing the SED modeling.
This is needed to recover as much as possible the parameter space in SFR and M$_*$.

\section{\label{conclusions}Conclusions}

In this pilot study, we have proposed to use a state-of-the-art statistical method using machine learning algorithm, the Approximate Bayesian Computation, to determine the best-suited SFH to be used to measure the physical properties of a subsample of COSMOS galaxies.
These galaxies have been selected in mass ($\log$M$_*>$8.5) and in redshift ($0.5<z<1$).
Furthermore, we impose that the galaxies should be detected in all UV-to-NIR bands with a SNR higher than 10.
We verified that these criteria do not bias the sSFR distribution of the sample.

To model these galaxies, we considered a smooth delayed-$\tau$ SFH  with or without a rapid and drastic change in the recent SFH, that is in the last few hundreds Myr.
We have built a mock galaxies SED using the SED fitting code CIGALE.
The mock SEDs have been integrated into the COSMOS set of broad band filters.
To avoid large dynamical ranges of fluxes which is to be avoided when using classification algorithms, we compute flux ratios.

Different classification algorithms have been tested with XGBoost providing the best results with a classification error of 20.98$\%$.
As output, the algorithm provides the probability that a galaxy is better modeled using a flexibility in the recent SFH.
The method is sensitive to variations of SFR that are larger than 1\,dex.

We have compared the results from the ABC new method with SED fitting using CIGALE.
Following the method proposed by \cite{Ciesla18}, we compare the results of two SED fits, one using the delayed-$\tau$ SFH and the other one adding a flexibility in the recent history of the galaxy.
The Bayesian Information Criterion are computed and compared to determine which SFH provides a better fit.
The BIC method provides a high error rate, 28$\%$, compared to the 21$\%$ obtained with the ABC method. 
Moreover, since the BIC method requires two SED fits per analyze of a source, it is much slower than the proposed ABC method: we were not able to compare them on the test catalog of $200,000$ sources and we had to introduce a smaller simulated catalog of size $30,000$ to compute their BIC in a reasonable amount time.

We use the result of the ABC method to determine the stellar mass and SFRs of the galaxies using the best-suited SFH for each of them.
We compare two samples of galaxies: the first one is galaxies with $\hat p(m=1 | \xobs)<$ 0.5, that are galaxies for which the smooth delayed-$\tau$ SFH is preferred, the second one is galaxies with $\hat p(m=1 | \xobs)>$0.75, galaxies for which there is a strong to decisive evidence against the smooth delayed-$\tau$ SFH.
The stellar mass distribution of these two samples is different.
The mass distribution of galaxies for which the delayed-$\tau$ SFH is preferred is similar to the distribution of the whole sample.
However, the mass distribution of galaxies needing a flexible SFH shows a deficit between 10$^{9.5}$ and 10$^{10.5}$\,M$_{\odot}$.
Their distribution is however similar to the whole sample's above M$_*=$10$^{10.5}$\,M$_{\odot}$.
Furthermore, the results of this study also implies that a flexible SFH is needed to cover the largest parameter space in terms of stellar mass and SFR possible, as seen from the sSFR distributions of galaxies with $\hat p(m=1 | \xobs)>$0.75.


\begin{acknowledgements}
The authors thank Denis Burgarella and Yannick Roehlly for fruitful discussions and the referee for his valuable comments that helped improve the paper.
The research leading to these results was partially financed via the PEPS Astro-Info program of the CNRS.
P. Pudlo warmly thanks the \textit{Centre International de Rencontres Math\'ematiques} (CIRM) of Aix-Marseille University for its support and the stimulating atmosphere during the Jean Morlet semester 'Bayesian modeling and Analysis of Big Data' chaired by Kerrie Mengersen. 

\end{acknowledgements}

\bibliographystyle{aa}
\bibliography{paper_ABC}
\appendix
\section{\label{apImp}Impact of fluxes SNR on the distribution of $p(x_{obs}|m=1)$}
\begin{figure}[!h] 
  	\includegraphics[width=\columnwidth]{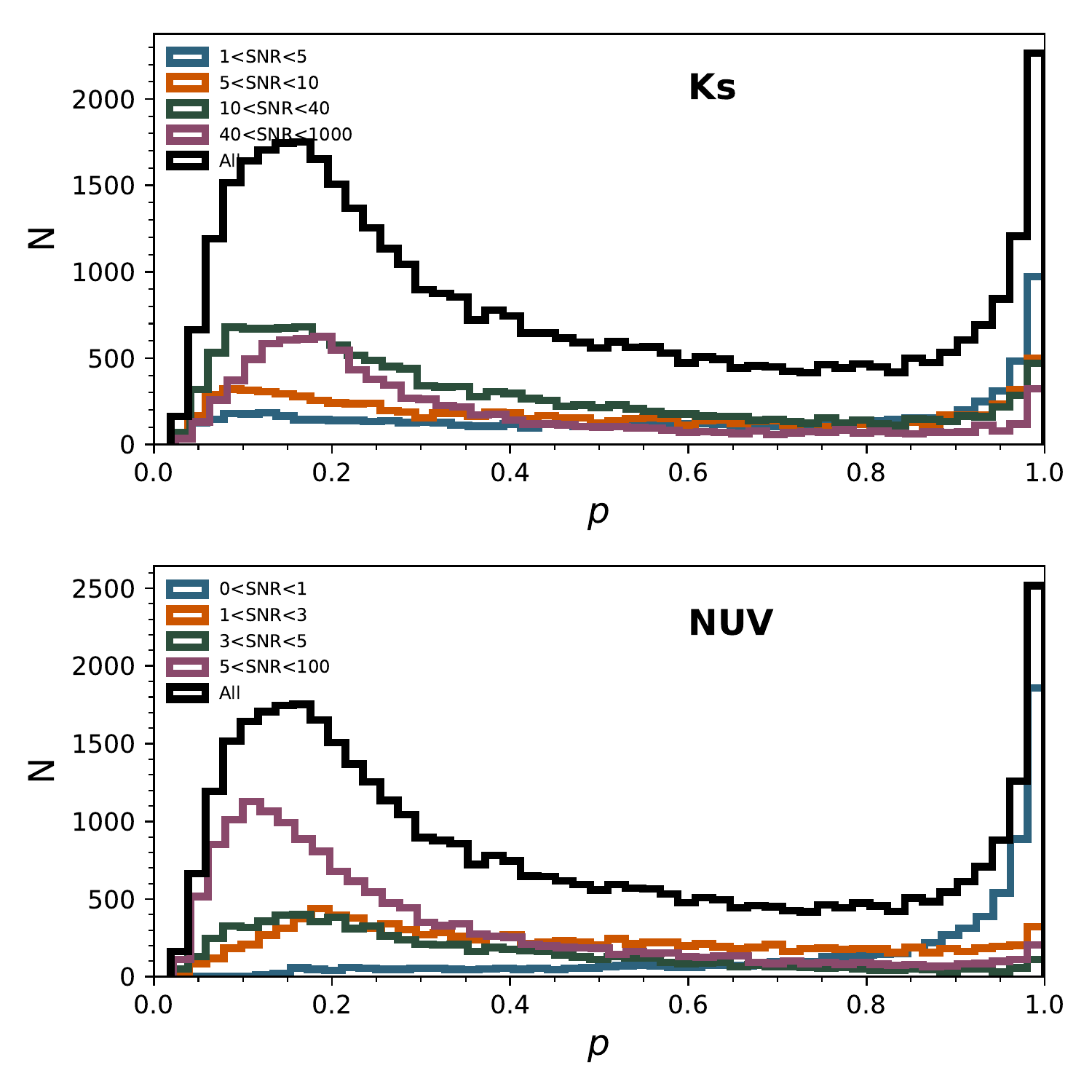}
  	\caption{Distribution of the predictions $\hat p(m=1 | \xobs)$  as a function of Ks band SNR (top panel) and NUV SNR (bottom panel). The different colors are for different selection in SNR in each panels. }
  	\label{pred_cosmos_SNR}
\end{figure}

In Fig.~\ref{pred_cosmos_SNR}, we show the distribution of the estimated probability $\hat p(m=1|x_{obs})$ for the subsample of COSMOS sources described in Sect.~\ref{sample} before applying any SNR cuts.
In this figure, all COSMOS sources with M$_*>$10$^{8.5}$\,M$_{\odot}$ and redshift between 0.5 and 1 are used.
The 0 value indicates that the delayed-$\tau$ SFH is preferred whereas $\hat p=1$ indicates that the delayed-$\tau$+flex SFH is more adapted to fit the SED of the galaxy.
To understand what drives the shape of the $\hat p(m=1 | \xobs)$ distribution, we show in the same figure the distributions obtained for different Ks SNR bins (top panel) and NUV SNR bins (bottom panel).
Galaxies with low SNR in either NUV and Ks photometric band show flatter $\hat p(m=1 | \xobs)$ distributions.
This means that these low SNR sources yields to intermediate values of $\hat p(m=1 | \xobs)$ , translating into a difficulty to choose between the delayed-$\tau$ and the delayed-$\tau$+flex SFHs.

\section{\label{apFit}Parameter tuning for Classification methods}
The training catalog is used to optimize the value of $\phi$ with a specific algorithm given $\psi$, and the validation catalog is used to fit the tuning parameters $\psi$.
To fit $\phi$ to a catalog of simulated datasets $\big(m^i, x^i\big)$, $i\in I$, the optimization algorithm specified with the machine learning model maximizes
\begin{multline*}
  \prod_{i\in I} L\Big(\hat p(m=1|x^i);\ {m^i}\Big) 
 \\
  \text{where }
 L(p;\ m) = \begin{cases}
     p & \text{if } m = 1, \\
     (1-p) & \text{if } m = 0,
 \end{cases}
\end{multline*}
given the value of $\psi$. Generally, this optimization algorithm is run for several values of $\psi$.  Then, the validation catalog is used to calibrate the tuning parameters $\psi$ based on data: the accuracy of $\hat p_\psi(m=1|x)$ for many possible values of $\psi$ is computed on the validation catalog and we select the value $\hat\psi$ that leads to the best results on this catalog. The resulting output of this two-step procedure is the approximation $\hat p_{\hat \psi}(m|x)$, that can be evaluated easily for new dataset $x'$. The accuracy of $\hat p(m=1|x)$ can be measured with various metrics. The most common metric is the classification error rate on a catalog of $\big(m^j, S(x^j)\big)$, $j\in J$, of $|J|$ simulations. We will rely on this metric. It is is defined by the frequency at which the datasets $x^j$ are not well classified, i.e.,
\begin{multline*}
    \frac{1}{|J|}\sum_{j\in J} \mathbf 1\Big\{ \hat m^j \neq m^j\Big\}, 
\text{where }
\hat m^j = \begin{cases}
    1 & \text{if }\hat p(m=1|x^j) > 1/2.
    \\
    0 & \text{if }\hat p(m=1|x^j) \le 1/2.
\end{cases}
\end{multline*}

\end{document}